\documentclass[a4paper,11pt]{article}
\usepackage{pos}
\usepackage{hyperref}
\usepackage{xcolor}
\usepackage{natbib} 
\usepackage{tcolorbox}
\usepackage{soul}

\newcommand{\greencheck}{{\color{green}\checkmark}}
\newcommand{\redcross}{{\color{red}$\times$}}
\newcommand{\fiftyfifty}{{\color{gray}(50/50)}}

\title{ChatQCD: Let Large Language Models Explore QCD}

\author*[a]{Antonin Sulc}
\author[b]{Patrick L.S. Connor}

\affiliation[a]{Helmholtz-Zentrum Berlin fuer Materialien und Energie,\\
Albert-Einstein-Str. 15, 12489 Berlin, Germany}

\affiliation[b]{Center for Data and Computing in Natural Sciences,\\
Universität Hamburg, Germany,\\
Albert-Einstein-Ring 10, D-22761 Hamburg}

\emailAdd{antonin.sulc@helmholtz-berlin.de}
\emailAdd{patrick.connor@desy.de}

\abstract{Quantum chromodynamics (QCD) has yielded a vast literature spanning distinct phenomena. We construct a corpus of papers and build a generative model. This model holds promise for accelerating the capability of scientists to consolidate their knowledge of QCD by the ability to generate and validate scientific works in the landscape of works related to QCD and similar problems in HEP. Furthermore, we discuss challenges and future directions of using large language models to integrate our scientific knowledge about QCD through the automated generation of explanatory scientific texts.}

\FullConference{%
  42nd International Conference on High Energy Physics\\
  17-24 July 2024\\
  Prague Congress Centre, Prague, Czech Republic
}

\begin{document}
\maketitle

\section{Introduction}
This paper introduces ChatQCD, an approach that utilizes large language models (LLMs) to explore and consolidate QCD knowledge from a comprehensive corpus of research papers. The motivation for this work is multifaceted: the complexity of QCD literature necessitates advanced tools for efficient knowledge synthesis; modern artificial intelligence techniques offer promising avenues for accelerating scientific understanding, and LLMs have the potential to enhance the accessibility of QCD research for both experts and newcomers. Furthermore, this approach may serve as a model for knowledge consolidation in other scientific disciplines by leveraging existing research pipelines. 

\section{Method}
\paragraph{Dataset.}
We compiled a dataset of 45,422 arXiv PDF documents, which were converted to an uniform text format using Nougat OCR~\cite{blecher2023nougat}. This tool efficiently processes common scientific paper formats, exporting formulas and tables in \LaTeX{}~and structuring text into sections.
The training dataset comprises two components: (1) an unsupervised dataset derived from raw documents, including tables and LaTeX formulas, and (2) a supervised dataset generated through a self-instruct~\cite{wang2022self}. It consists of question-answer pairs generated by \texttt{Mistral-7B-Instruct-v0.3} model by prompting paper paragraphs to 10 question-answer pairs in JSON format.

\paragraph{Model Architecture and Fine-tuning.}
We utilized the Unsloth Framework~\cite{unsloth2023}, a lightweight system optimized for efficient fine-tuning of LLMs. The base model is \texttt{llama-3-8b-Instruct-bnb-4bit}.
The model was fine-tuned using Low-Rank Adaptation (LoRA)~\cite{hu2021lora}, a parameter-efficient technique that adapts the model via matrix multiplication.
Key parameters for the LoRA configuration included a rank ($r$) of 32 and an $\alpha$ value of 8. Training arguments were configured as follows: per-device train batch size was set to $16$, gradient accumulation steps to $32$, with only $5$ warmup steps and $1$ training epoch. The learning rate ($\gamma$) was set to $2^{-4}$, using an 8-bit AdamW optimizer regularized by a weight decay $\lambda$ of $0.005$, with linear scheduling of the learning rate.

\newcommand{\qa}[4]{
\resizebox{1.0\linewidth}{!}{
	\begin{tcolorbox}
	\texttt{#1}
	\tcblower
	Claude 3.5: \texttt{#2}\\
	Base: \texttt{#3}\\
	ChatQCD: \texttt{#4}\\

\end{tcolorbox}}
}

\section{Experiments and Evaluations}
\resizebox{1.0\linewidth}{!}{
\begin{tabular}{cc}
\qa{Question (1): What's the QCD vacuum's $\theta$ parameter value?}
{\greencheck $\approx 0$}
{\greencheck 0.} 
{\greencheck The QCD vacuum's $\theta$ parameter value is zero.} &
\qa{Question (2): What's the $\beta$ function's leading coefficient in QCD?}
{\redcross $-11N_c/3$}
{\redcross 1.}
{\greencheck The beta function's leading coefficient in QCD is $b_{0}=\frac{1}{(4\pi)^{2}}\left(\frac{11}{3}N_{c}-\frac{2}{3}N_{f}\right)$}\\
\qa{Question (3): How does the QCD coupling constant change with energy scale?}
{\greencheck It decreases at higher energies due to asymptotic freedom.}
{\fiftyfifty The QCD coupling constant, $\alpha_s$, decreases with increasing energy scale (or decreasing momentum transfer) according to the renormalization group equation.}
{\greencheck The QCD coupling constant decreases with energy scale.} & 
\qa{Question (4): Why is QCD challenging to solve in the low-energy regime?}
{\greencheck Strong coupling makes perturbative methods ineffective at low energies.}
{\greencheck QCD is challenging to solve in the low-energy regime because the theory is non-perturbative, meaning that the calculations become increasingly difficult as the energy scale decreases.} 
{\fiftyfifty QCD is challenging to solve in low-energetic regime because it is a non-perturbative theory that involves strong interactions, which are difficult to handle using perturbation theory.}
\end{tabular}}

We analyze the logits, which serve as indicators of an LLM's confidence in predicting the next token in a sequence. We present two sequences where correct and incorrect statements are provided to the model, followed by a measurement of their respective logits (incorrect options are \st{striked}):
(1) \texttt{How many gluons exist? (8|\st{9})}
	The logit for the correct answer \texttt{8} was $\approx 3.53$, while for the incorrect \texttt{9}, it was $\approx 1.14$.
(2) \texttt{In QCD, quarks interact via the exchange of gluons, which carry (color|\st{mass}) charge.}
	The logit for the correct term \texttt{color} was $\approx 10.73$, whereas for the incorrect \texttt{mass}, it was $\approx 8.20$.
These results demonstrate that in both cases, the logits (i.e., the model's confidence) are higher for the correct statements.

\section{Conclusion}
ChatQCD demonstrates a promising LLM alternative to commercial models.
By utilizing a dataset of arXiv papers and employing advanced fine-tuning techniques, we have created a model capable of generating relevant responses to QCD-related queries, in some cases surpassing existing commercial models (Question 2), but qualify of the output depends on formulation of the question.
Our experiments show encouraging results regarding the model's confidence in predicting correct information.
However, challenges remain, particularly in handling complex scientific concepts and ensuring factual accuracy.

We propose the following improvements: (1) enhanced pre-selection of high-quality publications, (2) implementation of advanced feedback mechanisms~\cite{ouyang2022training,leerlaif},
(3) refinement of supervised instructions to eliminate misleading references, and (4) additional validation of source materials, as publication on arXiv does not guarantee content validity.

\paragraph{Acknowledgement:} We would like to thank to Radek Zlebcik for review of correctness of answers from our model.

\paragraph{Resources} Resources available at \href{https://github.com/sulcantonin/CHATQCD\_ICHEP24}{https://github.com/sulcantonin/CHATQCD\_ICHEP24}

\bibliographystyle{JHEP}
\bibliography{references.bib}

\end{document}